\documentstyle[12pt]{elsart}
\begin{document}

\def\beq{\begin{eqnarray}}
\def\eeq{\end{eqnarray}}
\begin{frontmatter}

\title{\sc A critique on the 
supplementary conditions of Rarita-Schwinger framework}

\author{\sc M. Kirchbach$^a$, D. V. Ahluwalia$^{a,b}$}

\address{$^a$ Theoretical Physics Group,  Facul. de Fisica\\
Univ. Aut. de Zacatecas,
Zacatecas, ZAC 98062, Mexico}

\address{$^b$Mail stop H-846,  Los Alamos National Laboratory\\ Los Alamos,
NM 87545, USA}

\begin{abstract}
After a brief review of the celebrated 1941 paper of Rarita and Schwinger 
on the theory of particles with half-integral spins, we present 
an {\em ab initio} construct of the representation space relevant 
for the description of spin-$\frac{3}{2}$
particles. The chosen example case of spin-$\frac{3}{2}$ shows that
covariance of a wave equation, and that of the imposed supplementary 
conditions, alone is not a sufficient criterion to guarantee 
the compatibility of a framework with relativity -- a lesson already 
arrived by Velo and Zwanziger. Here this same lesson is shown to be
true at the level of the representation space without invoking any
interactions.
The presented detailed analysis forces us to 
abandon the single-spin interpretation
of the   Rarita and Schwinger framework, and suggests a new interpretation
that fully respects the relativity theory.
\end{abstract}

\end{frontmatter}

\section{Introduction}

Higher spin fields have attracted attention since the very early
days of particle physics. After the introduction of the
neutrino into the theory of $\beta$-decay, Oppenheimer on the basis
of then-existing data put forward the suggestion that electron
neutrino may have a spin of three half,  and carry mass \cite{jo}. The 
problem of neutrino mass is being currently addressed 
experimentally,\footnote[1\label{f1}]{More precisely, what 
the current experiments 
are probing is the phenomena of flavor oscillations. 
In such a scenario a flavor eigenstate is a linear superposition
of different mass eigenstates\cite{bp1,bp2,mns,bp3,bbp}.} while the 
Oppenheimer's suggestion of a spin-$\frac{3}{2}$ neutrino was immediately
ruled out by a set of two papers \cite{rs,sk}. 

Though this month (July 2001) falls on the sixtieth anniversary of 
Ref. \cite{rs}, the 
interest in Rarita-Schwinger formalism remains unabated as more and more
baryonic resonances of higher spins are found in particle detectors on the one
hand, and as theorists realize that for one reason, or another, higher spins
may play a pivotal role in the unification of gravity with other interactions.
Yet, this sixty-year old formalism remains vexing to theorists to some
extent. This circumstance arises due to difficulties with the quantization
of Rarita-Schwinger fields on the one hand, 
and their tachyonic propagation on the other.  
We conjecture that the known difficulties associated with the
Rarita-Schwinger formalism may take their origin from the improper
treatment and interpretation of the underlying representation space. 
In conjunction with Ref. \cite{ak},
this paper is a preliminary step towards exploring this idea.

Here we first retrace the arguments of Rarita and Schwinger, and then 
immediately proceed to construct the representation space defined by Eq. 
(\ref{rsr}), below. This would allow us to present an essentially 
self-contained completion of the Rarita-Schwinger framework 
that is consistent with the relativity theory at the kinematic
level.  
At the same time it will allow us to point out where and how 
the inconsistency in the canonical wisdom on the  Rarita-Schwinger 
framework enters. 

Our considerations will be confined to spin-$\frac{3}{2}$. 
No new conceptual difficulties are expected to occur for 
spins $s>\frac{3}{2}$.

\section{Rarita-Schwinger framework for spin-$\mathbf\frac{3}{2}$}

The Rarita and Schwinger spinor-vector, $\psi_\mu$, transforms  as a  
finite dimensional non-unitary representation of the Lorentz group,
\beq
\psi_\mu:\quad
\underbrace{\left[\left(\frac{1}{2},0\right)\oplus\left(0,\frac{1}{2}\right)
\right]}_{\sc Spinor\;sector}
\otimes
\underbrace{\left(\frac{1}{2},\frac{1}{2}\right)}_{\sc Vector\; sector}
\label{rsr}
\eeq
In configuration space,
it satisfies the Dirac equation
\beq
\left(i\gamma^\lambda\partial_\lambda-m\right)\psi_\mu(x) =0\,,\label{rseq}
\eeq
for each of the Lorentz indices associated with 
the $\left(\frac{1}{2},\frac{1}{2}\right)$ representation space.
As is evident from Eq. (\ref{rsr}),
$\psi_\mu(x)$ contains $16$ degrees
of freedom. In the original interpretation, those were 
(correctly)interpreted to be distributed over two Dirac 
spinors, $\partial^\mu\psi_\mu(x)$ and $\gamma^\mu\psi_\mu(x)$,
and the eight degrees of freedom required for the description
of a spin-$\frac{3}{2}$ particle and its antiparticle.
The idea was put forward to nullify the indicated Dirac spinors in
the hope that in this way they will be removed from the representation 
space. In doing so one eventually would end up with eight 
degrees of freedom as required for the relativistic  description
of a spin-$\frac{3}{2}$ field.

That, $\psi_I(x)\equiv \partial^\mu\psi_\mu(x)$, and, $
\psi_{II}(x)\equiv\gamma^\mu\psi_\mu(x)$, indeed satisfy
the Dirac equation is immediately seen from the following two simple 
exercises:

\begin{enumerate}

\item[
{\bf I.}] 
Taking the divergence of Eq.~(\ref{rseq})
leads to
\beq
\partial^\mu (i\gamma^\lambda\partial_\lambda -m)\psi_\mu(x)&=& 
\left(i\gamma^\lambda\partial_\lambda -m\right)\partial^\mu\psi_\mu(x)=0\,,
\eeq
and therefore to
\beq
(i\gamma^\lambda\partial_\lambda -m)\psi_I(x)=0
\label{1st_DirSp}
\eeq
The situation with  $\psi_{II}(x)$ is
slightly trickier, though not fatally.

\item[{\bf II.}] 
In nullifying $\psi_I(x)$, i.e., in setting 
$\psi_I(x)=0$,  allows for
$\psi_{II}(x) $ to satisfy a Dirac  equation (with the wrong sign
for the mass term): 
\beq
(i\gamma^\lambda\partial_\lambda -m)\psi_{II}(x)
=
2i\psi_I(x) - \gamma^\mu(i\partial_\lambda\gamma^\lambda+m)\psi_\mu(x)
\eeq
The second term on the right-hand side of the
above equation carries a wrong sign for the mass term (if  $\psi_{II}(x)$
is to satisfy the Dirac equation). This, however, can be
corrected by replacing the 1941 Rarita-Schwinger  suggestion of $\psi_{II}$ 
by $\psi_{II}(x)^\prime\equiv \gamma^5\psi_{II}(x)$. 
Then, it is clear that
$\psi_{II}^\prime(x)$ satisfies the Dirac equation. This is an important point
as regards the relative intrinsic parities of the two 
spin-$\frac{1}{2}$ particles
contained in the representation space (\ref{rsr}). However, for the
rest of this paper we shall ignore this ``minor'' matter of inconsistency 
in the Rarita-Schwinger framework without
affecting our essential conclusions in any way. However, 
the reader should keep the presence of $\gamma^5$
in $\psi^\prime_{II}(x)$ in mind while applying the
framework to physical problems. 

\end{enumerate}

In summary, the Rarita-Schwinger framework for spin-$\frac{3}{2}$ consists
of Eq. (\ref{rseq}), supplemented by the conditions:

\beq
\gamma^\mu\psi_\mu(x)&=&0\,,\label{cond2}\\
\partial^\mu\psi_\mu(x) &=&0\,.\label{cond1}
\eeq

This framework is then claimed to describe a pure spin-$\frac{3}{2}$ system, 
despite a parenthetic remark in the original paper of Rarita and 
Schwinger which read, ``it [the square of the intrinsic 
angular momentum] will not have this value [$\frac{3}{2}$] in an arbitrary
reference frame.'' While our analysis will explicitly support this remark, 
we will show that despite covariance of the system of
Eqs. (\ref{rseq}), (\ref{cond2}), and (\ref{cond1}), the Rarita-Schwinger
framework is incompatible with the theory of relativity.

\section{Kinematic structure of the Rarita-Schwinger framework}

The most noted problems with the above summarized  
framework have been given by Johnson and Sudarshan \cite{js}, 
on one hand, and by Velo and Zwanziger 
\cite{vz} on the other. 
These authors studied propagation of Rarita-Schwinger field in an 
external electromagnetic potential.
In particular,  Velo and Zwanziger  
came to the conclusion that ``the 
main lesson to be drawn  $\ldots$ is that special relativity is not 
automatically satisfied by writing equations that transform covariantly.
In addition, the solutions must not propagate faster than light.''

Here, essentially the same is shown to be true purely at the level of 
the representation space without invoking any interactions --- provided
that one
takes due care, beyond the work of Refs. \cite{js,vz}, of the 
$\left(\frac{1}{2},
\frac{1}{2}\right)$ sector of the theory.
The detailed analysis presented in here 
forces us to abandon a single-spin interpretation
of the   Rarita and Schwinger framework, and suggests a new interpretation
that fully respects the relativity theory --- at least at the 
kinematic level.\footnote{We do not hasten to study as to what happens when
interactions are introduced. The reason is simple: if the kinematic structure
itself is acausal, or pathological in any manner, then these same 
elements would come to plague us later when interactions are introduced. 
In particular, we draw attention to Eq. (16) of Ref. \cite{ak} 
which indicates as
to what could have gone wrong even with the completeness relation
for the $\left(\frac{1}{2},
\frac{1}{2}\right)$ sector of the theory. For any theory that
does not satisfy the correct completeness relation, quantization
is bound to be problematic.}
The new 
interpretation of the representation space defined by Eq. (\ref{rsr})
will require us to abandon the  supplementary conditions,
(\ref{cond1}) and (\ref{cond2}), and force us to interpret this space as 
a multi-spin object containing two spin half objects of opposite relative
intrinsic parities, and a spin three-half object.

Our entire analysis, unless otherwise made apparent, 
will be done in the momentum space.

\subsection{Incompatibility of the Rarita-Schwinger framework with theory
of relativity}

The un-truncated Rarita-Schwinger representation space is a direct
product of a spinor and a Lorentz vector.
The objects which span the 
spinor and vector sectors of the theory are obtained by
applying the 
$\left(\frac{1}{2},0\right)\oplus\left(0,\frac{1}{2}\right)$ boost 
to the following rest, i.e.,$\vec p=\vec 0$, spinors:
\beq
\psi_1(\vec 0) = \left(
                  \begin{array}{c}
                   1\\
                   0\\
                   1\\
                   0
                   \end{array}
		\right)\,,
\psi_2(\vec 0) = \left(
                  \begin{array}{c}
                   0\\
                   1\\
                   0\\
                   1
                   \end{array}
		\right)\,,
\psi_3(\vec 0) = \left(
                  \begin{array}{c}
                   1\\
                   0\\
                   -1\\
                   0
                   \end{array}
		\right)\,,
\psi_4(\vec 0) = \left(
                  \begin{array}{c}
                   0\\
                   1\\
                   0\\
                   -1
                   \end{array}
		\right)\,,
\eeq
and by boosting 
the following 
$\left(\frac{1}{2},\frac{1}{2}\right)$
Lorentz  rest-frame vectors:
\beq
w_1(\vec 0) = \left(
                  \begin{array}{c}
                   1\\
                   0\\
                   0\\
                   0
                   \end{array}
		\right)\,,
w_2(\vec 0) = \frac{1}{\sqrt2}
                 \left(
                  \begin{array}{c}
                   0\\
                   1\\
                   1\\
                   0
                   \end{array}
		\right)\,,
w_3(\vec 0) = \left(
                  \begin{array}{c}
                   0\\
                   0\\
                   0\\
                   1
                   \end{array}
		\right)\,,
w_4(\vec 0) = \frac{1}{\sqrt2}
                 \left(
                  \begin{array}{c}
                   0\\
                   1\\
                   -1\\
                   0
                   \end{array}
		\right)\,.
\eeq
The $\left(\frac{1}{2},0\right)\oplus\left(0,\frac{1}{2}\right)$ and
the $\left(\frac{1}{2},\frac{1}{2}\right)$ boosts are in turn given
by,
\beq
\kappa^{\left(\frac{1}{2},0\right)\oplus\left(0,\frac{1}{2}\right)
}
=
\kappa^{\left(\frac{1}{2},0\right)}
\oplus
\kappa^{\left(0,\frac{1}{2}\right)}\,,\quad
\kappa^{\left(\frac{1}{2},\frac{1}{2}\right)}
=
\kappa^{\left(\frac{1}{2},0\right)} 
\otimes
\kappa^{\left(0,\frac{1}{2}\right)}\,,
\eeq
with
\beq
\kappa^{\left(\frac{1}{2},0\right)}&=&
\frac{1}{\sqrt{2m(E+m)}}\left[(E+m)I_2+\vec\sigma\cdot\vec p\,\right]\,,\\
\kappa^{\left(0,\frac{1}{2}\right)}&=&
\frac{1}{\sqrt{2m(E+m)}}\left[(E+m)I_2-\vec\sigma\cdot\vec p\,\right]\,.
\eeq
All notational details are those of Ref.~\cite{ak}.

A careful reader has perhaps already noted that the application
of the boost operators takes one from the original laboratory frame
to a boosted frame. However, as no inertial frame is a preferred frame,
the boosted objects should also exist in the original frame. It is
by this ``Wigner argument'' that the laboratory frame is populated with
spinors and vectors for all values of $\vec p$.

While all $\psi_i(\vec p\,)$, $i=1,2,3,4$, carry well-defined
spin, i.e. $s=\frac{1}{2}$, same is {\bf not} true for 
$w_\zeta(\vec p\,)$, $\zeta=1,2,3,4$. However, for $\vec p=\vec 0$,
the latter, for $\zeta=1,2,3$ are eigenstates of spin one, 
while the $\zeta=4$ case yields spin zero. The interested reader 
will find that this result is in accord with observation of Rarita
and Schwinger in the context of spin three half --- see, the parenthetic
remark after Eq. (2) of Ref.~\cite{rs}.  
Complementary  details on the $\left(\frac{1}{2},\frac{1}{2}\right)$
representation space can be found in Ref.~\cite{ak}.

After rotation by the matrix,
\beq
S=\frac{1}{\sqrt{2}} 
\left(
\begin{array}{cccc}
0 & i & -i & 0 \\
-i & 0 & 0 & i \\
1 & 0 & 0 & 1 \\
0 & i & i & 0
\end{array}
\right)\,,
\eeq
obtained in  Ref.~\cite{ak},
the $w_\zeta(\vec p\,)$, carry the usual (contravariant)Lorentz index. 
We denote
this $S$-rotated object by $\left[W_\zeta(\vec p\,)\right]^\mu$,
or simply as $W^\mu_\zeta(\vec p\,)$ 
In this language (and in
momentum space), the
$16$ objects that span the representation space defined in Eq. (\ref{rsr})
are obtained as:
\beq
\psi_{i\zeta}^\mu (\vec p\,) \equiv
\psi_i(\vec p\,)\otimes W^\mu_\zeta(\vec p\,)
\,.
\eeq
In order for $\gamma_\mu\psi_{i\zeta}^{\mu}(\vec p)$ to 
identically vanish for {\em all\/}
values of $i$ and $\zeta$, 
\beq
\left(E+m\right)^2-\vec p^{\,2} =0.
\eeq
Solving for $E$, 
\beq
E=-m\pm\sqrt{\vec p^{\,2}}.
\eeq
As such, 
the group velocity associated with the Rarita-Schwinger field
turns out to be:
\beq
\vec v_g \equiv \frac{\partial E}{\partial \vec p}= 1 \,\widehat p
\eeq
That is, implementing the supplementary condition
(\ref{cond2}) requires the group velocity associated with the Rarita Schwinger
field to be unity (i.e. equals the velocity of light). This value is 
{\em independent of mass\/} of the
Rarita-Schwinger field under consideration. Consequently, we conclude that 
the covariance of a set of equations alone
is not sufficient to warrant consistency with the theory of relativity.
One must further demand obtaining the correct dispersion relation.

\subsection{The truncation of the 
$\left(\frac{1}{2},\frac{1}{2}\right)$ sector
as the origin of difficulties of the 
Rarita-Schwinger framework}

The supplementary condition (\ref{cond2}) involves not only a 
summation over the Lorentz indices, but also involves a 
transformation on the relevant spinorial elements. 
In contrast, the supplementary condition (\ref{cond1})
sums out the Lorentz index, and without any further transformation
on the spinorial element sets it equal to zero.
It is therefore instructive to look at the Lorentz-index defining
 $\left(\frac{1}{2},\frac{1}{2}\right)$ representation space, 
to gain further insight in the representation space (\ref{rsr}).

A direct calculation of the divergency of each one of the
four Lorentz vectors $W_\zeta^\mu(\vec p\,)$, $\zeta =1,2,3,4$, leads to
\beq
p_\mu  W_\zeta ^\mu (\vec p\,)& = & 
c_\zeta (m^2 -p^2) =0 
\quad  \mbox{for} \quad \zeta =1,2,3\, ,
\nonumber\\
p_\mu W_4^\mu (\vec p) & = &{i\over m}p^2\, .
\label{1st_AxC}
\end{eqnarray}
Here, $c_1=i(p_x+i p_y)$, $c_2=-ip_z$, and $c_3=-i(p_x-i p_y)$.
As long as the first supplementary condition on $p_\mu\psi^\mu(\vec p\,) $
operates onto the Lorentz index only, the latter equations show
that it checks consistency with the mass-shell relation
$E^2-\vec{p}\, ^2=m^2$. For massive particles this condition
is fulfilled only for 
vectors $W^\mu_1(\vec p\,)$, $W^\mu_2(\vec p\,)$, and $W^\mu_3(\vec p\,)$, 
and is not satisfied
at all for the vector $W^\mu_4(\vec p\,)$. This calculation shows 
that  imposing the       
supplementary  condition (\ref{cond1}) onto the Rarita-Schwinger 
field restricts
the underlying four vectorial degrees of freedom to only three. 
Yet, as shown in
Ref.~\cite{ak}, the three vectors $W^\mu_1(\vec p\,)$, $W^\mu_2(\vec p\,)$ 
and $W^\mu_3(\vec p\,)$ are not
eigenstates of the squared angular momentum $\vec{J}\, ^2$ and do not lend
themselves to pure spin-1 states. Rather they are eigenstates of the
parity operator,  which in the considered representation space is nothing
but the matrix of the metric tensor $ \eta_{\mu\nu}$=diag$(1,-1,-1,-1)$.
Consequently, though this supplementary condition restricts the four degrees
of freedom of the $\left({1\over 2},{1\over 2}\right)$ representation space
to only three, it does not restrict the spin degrees of freedom to
spin-1 only.\footnote{Moreover,
these three Lorentz vectors cannot span the 
$\left({1\over 2},{1\over 2} \right)$ space in the same mathematical sense 
as do the four Dirac spinors in the  
 $\left({1\over 2},0\right)\oplus
\left(0,{1\over 2}\right)$ representation space. It was shown in
Ref. \cite{ak} that this serious drawback is immediately rectified
by incorporating, $W^\mu_4(\vec p\,)$, the fourth natural companion 
of the three $W^\mu_1(\vec p\,)$, $W^\mu_2(\vec p\,)$ and $W^\mu_3(\vec p\,)$. 
By doing so, the 
$\left({1\over 2},{1\over 2} \right)$ representation space,
in exact parallel of the Dirac's
$\left({1\over 2},0\right)\oplus
\left(0,{1\over 2}\right)$ representation space, carries objects that have
positive as well as negative norm {\em before\/} quantization. In addition,
both of these spaces now become endowed with covariantly separated sectors
of definite parities.
We expect this new parallelism to
circumvent the difficulties of quantization pointed out by Weinberg 
\cite{sw1964}.}

The spin-0 piece is still there and mixes up with
spin-1 within $W_\zeta (\vec{p}\, )$ (for $\zeta =1,2,3$).
The immediate consequence of the covariant inseparability of the 
$\left({1\over 2},{1\over 2} \right)$ space into
spin-0 and spin-1 is the covariant inseparability of the
Rarita-Schwinger  field into a spin-$\frac{3}{2}$ and two 
spin-$\frac{1}{2}$ components.

Only within the rest frame, or in the helicity basis, 
does the separation between spin-0 and spin-1 take place \cite{ak}.

The essential additional physics lies in the fact that the Proca 
equation
\beq
\partial_\mu F^{\mu\nu} + m^2 A^\nu =0
\eeq
by construction satisfies, $\partial_\nu A^\nu=0$ (for $m\ne 0$).
However, as shown in Ref. \cite{ak}, ``$\partial_\nu A^\nu=0$''
cannot be satisfied for all relevant degrees of freedom 
in the massive 
$\left(\frac{1}{2},\frac{1}{2}\right)$ representation space without
violating the completeness relation. While the wave equation satisfied
by the $\left(\frac{1}{2},\frac{1}{2}\right)$ spanning $W^\mu_\zeta(\vec p\,)$,
contains all solutions of the Proca equation the converse is not true. 
The wave equation for  $W^\mu_\zeta(\vec p\,)$, which carry with
them a completeness relation exactly paralleling
the Dirac construct for spin-$\frac{1}{2}$,  is  \cite{ak}:
\beq
\left(\Lambda_{\mu\nu}p^\mu p^\nu - \epsilon \,m^2 I_4\right)_{\alpha\beta}
 W^\beta_\zeta(\vec p\,) = 0\, ,
\eeq
where $\epsilon$ equals $+1$ for $\zeta=4$ and is $-1$ for
$\zeta=1,2,3$. The $\Lambda_{\mu\nu}$ matrices are:
\beq
\Lambda_{00}&=&
\left(\begin{array}{cccc}
1 & 0 & 0 & 0 \\
0 & -1 & 0 & 0 \\
0 & 0 & -1 & 0\\
0 & 0 & 0 & -1
\end{array}\right),\,\,
\Lambda_{11}=
\left(\begin{array}{cccc}
1 & 0 & 0 & 0 \\
0 & -1 & 0 & 0 \\
0 & 0 & 1 & 0\\
0 & 0 & 0 & 1
\end{array}\right),\,\,
\Lambda_{22}=
\left(\begin{array}{cccc}
1 & 0 & 0 & 0 \\
0 & 1 & 0 & 0 \\
0 & 0 & -1 & 0\\
0 & 0 & 0 & 1
\end{array}\right), \nonumber\\
\Lambda_{33}&=&
\left(\begin{array}{cccc}
1 & 0 & 0 & 0 \\
0 & 1 & 0 & 0 \\
0 & 0 & 1 & 0\\
0 & 0 & 0 & -1
\end{array}\right),\,\,
\Lambda_{01}=
\left(\begin{array}{cccc}
0 & -1 & 0 & 0 \\
1 & 0 & 0 & 0 \\
0 & 0 & 0 & 0\\
0 & 0 & 0 & 0
\end{array}\right),\,\,
\Lambda_{02}=
\left(\begin{array}{cccc}
0 & 0 & -1 & 0 \\
0 & 0 & 0 & 0 \\
1 & 0 & 0 & 0\\
0 & 0 & 0 & 0
\end{array}\right), \nonumber\\
\Lambda_{03}&=&
\left(\begin{array}{cccc}
0 & 0 & 0 & -1 \\
0 & 0 & 0 & 0 \\
0 & 0 & 0 & 0\\
1 & 0 & 0 & 0
\end{array}\right), \,\,
\Lambda_{12}=
\left(\begin{array}{cccc}
0 & 0 & 0 & 0 \\
0 & 0 & -1 & 0 \\
0 & -1 & 0 & 0\\
0 & 0 & 0 & 0
\end{array}\right), \,\,
\Lambda_{13}=
\left(\begin{array}{cccc}
0 & 0 & 0 & 0 \\
0 & 0 & 0 & -1 \\
0 & 0 & 0 & 0\\
0 & -1 & 0 & 0
\end{array}\right)\,,\nonumber\\
\Lambda_{23}&=&
\left(\begin{array}{cccc}
0 & 0 & 0 & 0 \\
0 & 0 & 0 & 0 \\
0 & 0 & 0 & -1\\
0 & 0 & -1 & 0
\end{array}\right)
\eeq
The remaining $\Lambda_{\mu\nu}$ are obtained from the above
expressions by noting: $ \Lambda_{\mu\nu} = 
\Lambda_{\nu\mu}$.

For this reason, and several others
given in Ref. \cite{ak}, 
the Proca equation is not endowed with the 
complete physical content of the massive 
$\left(\frac{1}{2},\frac{1}{2}\right)$ representation space. 
In correcting the Rarita-Schwinger framework for the above
incompleteness one is forced to drop the supplementary conditions and 
required to reinterpret the $\psi_\mu$ as a multi-spin(multi-parity) object.

\subsection{Multi-spin character of $\psi_\mu$ from the
perspective of the Pauli-Lubanski vector}

Here we confirm the multi-spin valued nature of $\psi_\mu$ from a 
slightly different perspective but the equation of motion.
We use instead the properties of the squared-length of the 
Pauli-Lubanski vector.
We shall make our analysis in four steps. In the first step
we simply recall the definition of the Pauli-Lubanski vector, and that of 
the associated Casimir invariant, $C_2$. In the second step we consider
$C_2$ for the $(j,0)\oplus(0,j)$ representation space. The third step does
the same for the $(j,j)$ representation spaces by considering the special
case of $j=\frac{1}{2}$. Finally, the fourth step is devoted to
$\psi_\mu(\vec p)$.

{\bf I.} The Pauli-Lubanski vector is defined as
\beq
{\cal W^\mu}={1\over 2}\epsilon^{\mu \nu\rho\sigma }I_{\nu \rho}P_\sigma
\,.\label{P_L}
\eeq
Here, $\epsilon_{\mu\nu\rho\sigma}$ is the standard Levi-Cevita
symbol in four dimensions, while $I_{\nu\rho}$ denote the generators
of the Lorentz group
\beq
I_{0i}= \kappa_i\, , &\qquad & I_{ij}= \epsilon_{ijk}J^k\, , 
\eeq
with $\kappa_i$ and $J_k$ being in turn the $i$th and $k$th components
of boost and rotation generators, respectively.
The final expression for the Pauli-Lubanski vector in terms of
boost and rotation generators reads 
\beq
{\cal W}^\mu = \left(-\vec J\cdot \vec P,\;\;
 -\vec J P_0 + \vec \kappa\times\vec P\right)\,.
\label{PL_KJ}
\eeq
Correspondingly, its squared length 
is obtained as
\beq
C_2 \equiv  {\cal W}^\mu{\cal W}_\mu &=&
\left(\vec J\cdot\vec P\right)^2 - 
\left(-\vec J P_0 +\vec \kappa\times\vec P\right)^2\,. 
\label{PL_sq}
\eeq
So far this result is entirely general, i.e., it is true for an arbitrary
representation space of the Lorentz group.

{\bf II.} The detailed physical content
of $C_2$ is determined by the representation space under consideration.
The most straight-forward interpretation of $C_2$ occurs for
the $(j,0)$ and $(0,j)$ representation spaces. For these spaces the 
{\em boost\/} 
generators  are entirely determined by the $(2j+1)\times(2j+1)$ matrices
associated with the generators of {\em rotations\/},
\beq
\vec\kappa^{(j,0)}= + i\vec J\,,\quad 
\vec\kappa^{(0,j)}- i\vec J\,.\label{j}
\eeq
Consequently, the $C_2$'s take the simple form
\beq
C_2^{(j,0)} = C_2^{(0,j)}=\left(\vec J\cdot\vec P\right)^2 
- \left(\vec J P_0\right)^2 + \left(\vec J\times\vec P\right)^2
\eeq
In order to obtain a parity covariant framework 
one constructs $(j,0)\oplus(0,j)$ representation spaces. It was
shown in Refs. \cite{bww,ijmpe} that 
each of these  representation spaces is $2(2j+1)$ dimensional,
is endowed with a covariant parity bifurcation into $(2j+1)$
subspaces of {\em opposite} relative intrinsic parities, and supports
particles and antiparticles of spin-$j$. 
For $j=\frac{1}{2}$, one obtains the standard Dirac representation space
-- provided one seeks objects that are eigenstates of the charge 
operator. Under similar assumption, for $j=1$ one obtains the 
Bargmann-Wightman-Wigner (BWW) representation space. Since 
\beq
\vec J^{\,\left(\frac{1}{2},0\right)\oplus\left(0,\frac{1}{2}\right)}
=\left(\begin{array}{cc}
{\vec\sigma}/{2} & 0 \\
0 & {\vec\sigma}/{2}
\end{array}\right)\,,\quad
\vec J^{\,\left(1,0\right)\oplus\left(0,1\right)}
= \left(\begin{array}{cc}
\vec S & 0\\
0 &\vec S
\end{array}\right)\,.
\eeq
where $\vec \sigma$ are the usual Pauli matrices, and $\vec S$ are
$3\times 3$ spin-1 matrices, it readily verifies that
\beq
C_2^{\,\left(\frac{1}{2},0\right)\oplus\left(0,\frac{1}{2}\right)}
\;\psi_i^{\,\left(\frac{1}{2},0\right)\oplus\left(0,\frac{1}{2}\right)}
(\vec p\,)
= -\frac{3}{4} m^2\; 
\psi_i^{\,\left(\frac{1}{2},0\right)\oplus\left(0,\frac{1}{2}\right)} 
(\vec p\,)\,,
\\
C_2^{\,\left(1,0\right)\oplus\left(0,1\right)}\;
\psi_i^{\,\left(1,0\right)\oplus\left(0,1\right)}(\vec p\,)
= -\,2\,m^2\;
\psi_i^{\,\left(1,0\right)\oplus\left(0,1\right)}(\vec p\,)
 \,,
\eeq
where $i=1,2,3,4$ for the Dirac spinors, and 
$i=1,2,3,4,5,6$ for the BWW objects.
Thus, we explicitly
verify that these representation spaces
are correctly associated with pure spin one half, and one, respectively:

It is useful to summarize the  following observations
for the  $(j,0)\oplus(0,j)$ representation space:
\begin{enumerate}
\item
The eigenstates of the $C_2^{(j,0)\oplus(0,j)}$'s are also eigenstates of 
$\left(\vec J^{(j,0)\oplus(0,j)}\right)^2$. Mathematically,
this happens because commutator of $C_2$ and $\vec J^{\,2}$ vanishes,
\beq
\left[C_2,\;\vec J^{\,2}\right]_{(j,0)\oplus(0,j)}
=0\,.\label{c1}
\eeq

\item
The $2(2j+1)$ dimensional space bifurcates into two sectors 
of opposite relative intrinsic parties. Each of these sectors
carries dimensionality $2j+1$, and are related to each other by the
action of a charge conjugation operator.
\end{enumerate}

{\bf III.}
For other representation spaces no simple relationship, such
as given by Eqs. (\ref{j}), exists between
the generators of boosts and the generators of rotations.
For this reason the result just summarized for the 
$(j,0)\oplus(0,j)$ space  no longer remains true.

Specifically, consider the $\left(\frac{1}{2},\frac{1}{2}\right)$
representation space. It is spanned by the four $W^\mu_\zeta(\vec p)$,
$\zeta=1,2,3,4$. The $C_2^{\left(\frac{1}{2},\frac{1}{2}\right)}$
is obtained by substituting the 
${\left(\frac{1}{2},\frac{1}{2}\right)}$ generators of rotations and boosts
into Eq. (\ref{PL_sq}).
These, for the specific realization of the  $W^\mu_\zeta(\vec p)$, are
the $S$-transformed generators given in Eqs. (4), (5), (6), and 
(7) of Ref. \cite{ak}. 
A detailed calculation then  shows that,
\beq
C_2^{\left(\frac{1}{2},\frac{1}{2}\right)}
 W^\mu_\zeta(\vec p\,) &=& - m^2 \lambda_\zeta W^\mu_\zeta(\vec p\,)\,,\quad
\left({\mbox{no sum on}}\,\,\zeta\right)\,.
\eeq
where $\lambda_\zeta=2$, for $\zeta=1,2,3$; and  
$\lambda_\zeta=0$ for $\zeta=4$. 

To avoid confusion, we note that $\lambda_\zeta$ can also be  
read off from Eq. (\ref{PL_sq}) by going to the rest frame.
The $C_2$, in general, while acting upon a mass eigenstate yields,
$- m^2\vec{J}\, ^2$. Because of that the eigenvalues of the
squared-length of the Pauli-Lubanski vector, i.e. $C_2$, 
at rest can be given the interpretation
of, $-m^2\lambda_\zeta =  - m^2j_\zeta(j_\zeta+1)$. 
Based upon this finding, valid solely at rest, the
impression arises that the 
eigenstates of the 
second Casimir of the Poincar\'e group
carry definite mass and a pure spin.
This impression, as regards the spin,
is in general not correct.\footnote{Further,
in reference to footnote [\ref{f1}], 
quantum framework allows for linear superposition of mass 
eigenstates. This  has important implications not only for neutrino
oscillation  phenomenology but it also has important
significance for the meaning of
gravitational potential in the quantum framework\cite{plb2000}.}
Indeed, while the eigenvalues of $C_2 $, in being a Casimir operator,
are frame independent, their association with the eigenvalues
of $\vec{J}^2$ {\it is frame dependent\/}.
In the most general case the eigenvalues of $C_2 $ arise as a consequence
of a delicate cancellation between the
actions of all the terms on the right hand side  of Eq.~(\ref{PL_KJ}) 
upon the state vectors. As a result of these cancellations, even
though the eigenvalues of $C_2 $ numerically coincide with the eigenvalues
of, $ - m^2 \vec{J}\, ^2 $, at rest, it has not to be confused with the latter.
Only for the $(j,0)\oplus(0,j)$ representation
space, the $\vec J^{\,2}$ and $C_2 $ have simultaneous eigen-basis (in all
inertial frames).

To summarize,

\begin{enumerate}
\item
The eigenstates of the 
$C_2^{\left(\frac{1}{2},\frac{1}{2}\right)}$'s are, in general,
not eigenstates of 
$-m^2\,\left(
{\vec J}^{\,\left(\frac{1}{2},\frac{1}{2}\right)} \right)^2$. Mathematically,
it derives it origin from the fact that the following commutator does not
vanish,
\beq
\left[ C_2,\;\vec J^{\,2}\right]_{\left(\frac{1}{2},\frac{1}{2}\right)}
\ne 0\,.\label{c2}
\eeq

\item
The $4$ dimensional space bifurcates into two sectors 
of opposite relative intrinsic parties. The one sector has
dimensionality of $3$, while the second sector has dimensionality
of unity. 
The charge conjugate sector is obtained by complex 
conjugating the $W^\mu(\vec p)$.
\end{enumerate}

 {\bf IV.}
 For representation spaces of the
Rarita-Schwinger type even though the action of $C_2$,
associated with the representation space defined by Eq. (\ref{rsr}), on mass
eigenstates $\psi_\mu(\vec p)$ formally yields, $-m^2 \lambda$,
with $\lambda=\frac{3}{4}$, twice, and $\lambda=\frac{15}{4}$, once,
 such states, in general, are not
eigenstates of $\vec J^{\,2}$ --- 
where $\vec J$ is the appropriate generator of
rotation. For the representation space defined by Eq. (\ref{rsr}),
it is possible, in a rest frame, say, to identify the eigenstates of
$\vec J^{\,2}$ 
as two objects of spin one half, and an object of spin three half.
However, this separation into  spin-states 
is not covariant in general --- even though
the associated $C_2$ divides the space (\ref{rsr}) into sub-spaces 
that carry eigenvalues, $-m^2\,j(j+1)$, with $j=1/2,1/2,3/2$. The latter 
eigenvalues of $C_2$ no longer carry meaning of ``spin'' in the sense
of being eigenstates of $\vec J^{\,2}$ (which they are not). 
This interpretation is in accord with the conjecture that one of us
advanced some years ago while studying the baryonic spectra \cite{k1,k2}.
It arises as a consequence of Eqs. (\ref{c1}) and (\ref{c2}), which imply,
\beq
\left[C_2,\;\vec J^{\,2}\right]_{\left[\left(\frac{1}{2},0\right)
\oplus\left(0,\frac{1}{2}\right)\right]\otimes\left(\frac{1}{2},
\frac{1}{2}\right)} \ne 0\,.
\eeq
on the one hand, and on the other hand due to Eqs. (\ref{cond1}) and 
(\ref{cond2}), which despite their covariance, are inadmissible on grounds
of being inconsistent with the theory of relativity.

\section{Conclusion}


Covariance of a set of equations alone does not guarantee their
compatibility with the theory of relativity. Specifically,
the supplementary condition
(\ref{cond2}) forces the {\em massive\/} Rarita-Schwinger field to have 
a group velocity of unity (clearly in violation of the theory
of relativity). The supplementary condition (\ref{cond1})
cannot be implemented for
{\em all\/} four degrees of freedom that span the massive $\left(\frac{1}{2},
\frac{1}{2}\right)$ representation space --- the failure arises
with $W^\mu_4(\vec p\,)$.

Once a charge conjugate part of a representation is incorporated
in the framework,\footnote{Such a charge conjugate
sector, e.g., is already present in the 
$\left(\frac{1}{2},0\right)\oplus\left(0,\frac{1}{2}\right)$ part of
the Rarita-Schwinger field.
For the $\left(\frac{1}{2},\frac{1}{2}\right)$ representation space
it can be shown to be brought in by the operation of complex conjugation.}
 the eigenvalues of the Casimir invariant $C_2$
split the representation space (\ref{rsr}) 
into $2(2j+1)$ dimensional subspaces,
with $j=\frac{1}{2}$ twice, and  $j=\frac{3}{2}$ once.

These subspaces further subdivide into sectors of definite 
relative intrinsic parities. In general, however, these subspaces, 
do  not carry a definite spin. That is, they are not eigenstates
of the square of the relevant generators of the rotations, $\vec J^{\,2}$.
The result readily extends to Rarita-Schwinger fields of spin greater
than three half. 

Only in the $\left(j,0\right)\oplus\left(0,j\right)$ representation spaces 
do these subspaces carry definite values of  
$\vec J^{\,2}$, $C_2$, and the parity operator as well.


We are thus left with no choice but to abandon a single-spin interpretation
of the representation space defined in Eq. (\ref{rsr}). Once that is done
$\psi_\mu$ contains two spin one half particles of opposite relative parities,
and a spin three half particle. In the absence of any interaction these
particles are mass degenerate. Furthermore, the bifurcation into 
spin one half
and spin three half occurs only in the rest and helicity frames. In general,
the particles in this representation do not carry a well-defined spin. 
In this representation space the Dirac 
operator $\left(\gamma^\lambda p_\lambda 
\pm m
\right)$
annihilates the spinorial sector of the $\psi_\mu(\vec p\,)$, while a new 
operator, $\left(\Lambda_{\lambda\nu} p^\lambda p^\nu
\pm m^2 I_4\right)$, annihilates
the vector sector of $\psi_\mu(\vec p\,)$.

\section*{Acknowledgments}
This work was was done in part under the auspices of the U.S.
Department of
Energy, and in part supported by CONACyT, Mexico. In addition,
we thank the Theory and Physics Divisions of the Los Alamos Laboratory
for providing the facilities for this work during the Summer of  2001

\end{document}